\documentclass[sigconf]{acmart}

\AtBeginDocument{%
  }

\usepackage{listings}
\usepackage{xcolor}
\usepackage{natbib}

\definecolor{customgray}{HTML}{404040}

\definecolor{codegreen}{rgb}{0,0.6,0}
\definecolor{codegray}{rgb}{0.5,0.5,0.5}
\definecolor{codepurple}{rgb}{0.58,0,0.82}
\definecolor{backcolour}{rgb}{0.95,0.95,0.92}
\lstdefinestyle{code-style}{
    backgroundcolor=\color{backcolour},   
    commentstyle=\color{codegreen},
    keywordstyle=\color{magenta},
    numberstyle=\tiny\color{codegray},
    stringstyle=\color{codepurple},
    basicstyle=\ttfamily\footnotesize,
    breakatwhitespace=false,         
    breaklines=true,                 
    captionpos=b,                    
    keepspaces=true,                 
    numbers=left,                    
    numbersep=5pt,                  
    showspaces=false,                
    showstringspaces=false,
    showtabs=false,                  
    tabsize=2
}

\definecolor{lightergray}{RGB}{240,240,240}

\begin{document}

\title{Using Copilot Agent Mode to Automate Library Migration: A Quantitative Assessment}

\author{Aylton Almeida}
\orcid{0009-0002-0649-856X}
\email{ayltonalmeida@dcc.ufmg.br}
\affiliation{%
  \institution{Federal University of Minas Gerais}
  \city{Belo Horizonte}
  \state{Minas Gerais}
  \country{Brazil}
}

\author{Laerte Xavier}
\orcid{0000-0001-7925-4115}
\email{laertexavier@pucminas.br}
\affiliation{%
  \institution{Pontificial University of Minas Gerais}
  \city{Belo Horizonte}
  \state{Minas Gerais}
  \country{Brazil}
}

\author{Marco Tulio Valente}
\orcid{0000-0002-8180-7548}
\email{mtov@dcc.ufmg.br}
\affiliation{%
  \institution{Federal University of Minas Gerais}
  \city{Belo Horizonte}
  \state{Minas Gerais}
  \country{Brazil}
}

\begin{abstract}
Keeping software systems up to date is essential to avoid technical debt, security vulnerabilities, and the rigidity typical of legacy systems. However, updating libraries and frameworks remains a time-consuming and error-prone process. Recent advances in Large Language Models (LLMs) and agentic coding systems offer new opportunities for automating such maintenance tasks. In this paper, we evaluate the update of a well-known Python library, SQLAlchemy, across a dataset of ten client applications. For this task, we use the Github's Copilot Agent Mode, an autonomous AI system capable of planning and executing multi-step migration workflows. To assess the effectiveness of the automated migration, we also introduce Migration Coverage, a metric that quantifies the proportion of API usage points correctly migrated. The results of our study show that the LLM agent was capable of migrating functionalities and API usages between SQLAlchemy versions (migration coverage: 100\%, median), but failed to maintain the application functionality, leading to a low test-pass rate (39.75\%, median).

\end{abstract}


\maketitle

\section{Introduction}

A key software engineering concern is keeping applications up to date to prevent them from becoming rigid legacy systems that are costly and difficult to maintain. Legacy systems often hinder innovation, increase technical debt, and expose organizations to security and compatibility risks. A critical part of this process is the continuous update of libraries and frameworks, which frequently evolve to provide new features, fix vulnerabilities, and improve performance~\cite{hou2011exploring,lamothe2021systematic,salama2019stability,jss-2018-deprecation}. When these dependencies are not updated regularly, client applications can quickly become obsolete, making future migrations more complex and expensive.

Recently, Large Language Models (LLMs) have emerged as powerful tools capable of automating a wide range of software engineering tasks, including those related to the migration and update of libraries and frameworks~\cite{migrating-code-at-scale-google,nikolov2025googleusingaiinternal,islam2025empiricalstudypythonlibrary}. In a previous study, we conducted an initial investigation on the update of a specific library (SQLAlchemy) within a single client project~\cite{automatic-library-migration-using-llms}. For that experiment, however, we relied on GPT-4's API  and explored three types of prompts: zero-shot, few-shot, and chain-of-thought. The results were promising but still far from achieving full automation of the migration process. For example, the LLM made minor mistakes, such as importing entities from incorrect classes and producing low-quality code that does not follow best practices enforced by well-known Python linters.

In this paper, we present an extension of our previous study. Essentially, we improved our previous work in three key dimensions:

\begin{itemize}

\item {\em Dataset:} Instead of evaluating the migration of SQLAlchemy using a single client project, we carefully built a dataset of ten client applications that use this library. All of them compile successfully, include fully passing test suites, and can be executed end-to-end.

\item {\em Agentic Approach}: We performed and evaluated the migration using an agentic system, specifically Github's Copilot Agent Mode.\footnote{https://code.visualstudio.com/docs/copilot/chat/chat-agent-mode} An agentic system is an AI-based development environment that can coordinate and execute complex programming tasks. Unlike chat-based AI tools, these agents can plan, reason, and perform multi-step workflows, including updating libraries and frameworks, without requiring constant human supervision~\cite{practical-considerations-for-agenti-llms-systems,agentic-llms-survey}. In our research, we decided to use GitHub’s Copilot Agent Mode. Copilot is one of the most widely adopted AI developer tools, with over 1.3 million paid subscribers and deep integration into common IDEs such as Visual Studio Code and the JetBrains suite~\cite{cio-copilot-2024-usage}. Recently, it also incorporated an agent mode, labeling it as ``the next evolution in AI-assisted coding''~\cite{microsoft-copilot-agent-mode}.

\item {\em Quantitative Assessment:} To evaluate our approach, we proposed and used a novel metric called Migration Coverage. Inspired by traditional test coverage metrics~\cite{characterizing-python-library-migrations}, Migration Coverage measures the percentage of call sites  in the codebase that were correctly migrated to a new version of a given API.

\end{itemize}

The remainder of this paper is organized as follows. Section~\ref{sec:study-design} describes the methodology used to build our  dataset and migrate it to SQLAlchemy v2. Section~\ref{sec:results} presents the migration results, including metric performance and comparisons with our previous work. Section~\ref{sec:threats-to-validity} discusses threats to validity, and Section~\ref{sec:related-work} reviews related studies. Finally, Section~\ref{sec:conclusion} concludes the paper.

\section{Study Design}
\label{sec:study-design}
To investigate the effectiveness of AI-based agents in supporting API migration, we conduct a study focused on upgrading real-world Python applications. This section details our methodology, beginning with the selection of our target API and the creation of a dataset of client applications. We then describe the migration process using Github's Copilot Agent Mode and conclude with the set of metrics used to evaluate the correctness and quality of the migration results.

\subsection{Target API: SQLAlchemy}
SQLAlchemy\footnote{https://GitHub.com/sqlalchemy/sqlalchemy} is an Object-Relational Mapping (ORM) tool that facilitates communication between Python applications and relational databases. It simplifies development by abstracting complex database connection and data manipulation tasks. 
In this paper, we address the migration from SQLAlchemy version 1 to version 2. While version 1 is widely-adopted, version 2 was introduced to leverage recent advancements in the Python ecosystem. Among the most significant changes is the full integration with Python's static typing, a feature that enhances error detection during development and eases the maintenance of larger codebases. Furthermore, version 2.0 offers major performance improvements and optimizations for asynchronous operations.

\subsection{Dataset Creation}

To gather a diverse set of client applications, we created a dataset of repositories that use the SQLAlchemy library. This was done using a crawling script that interacted with GitHub’s GraphQL API. The script  
fetched repositories that explicitly mentioned the library. Only repositories with at least 50 stars and created since 2018 were included to ensure a minimum level of community interest and relevance. 
Following the automated collection, a manual curation process was applied to select repositories suitable for the experiment. Repositories without passing tests were removed, as well as repositories that already used SQLAlchemy version 2. Lastly, only projects where the application and tests could be executed successfully were retained.
The initial query yielded 135 candidate repositories. From this total, we discarded 44 for lacking passing tests, 49 for having already been migrated, and 32 due to execution failures. This cleaning step resulted in a final dataset of 10 repositories for our experiment.

\subsection{Migration Process}
The migration for each repository was performed in two steps. The first was the creation of a GitHub Copilot instructions file.\footnote{https://docs.github.com/pt/copilot/how-tos/configure-custom-instructions/add-repository-instructions?tool=visualstudio} This file provides the LLM specific instructions, outlining the necessary code modifications for upgrading SQLAlchemy from version 1.x to 2.x. The second step was the migration itself, performed by issuing a prompt to Copilot's Agent Mode, which utilizes the GPT-4o model. 

To mitigate threats to validity, we also made two key decisions. First, the migration prompt for each project was executed only once to avoid variability in the LLM's output. Second, whenever the agent requested input or clarification, we provided the standardized response "keep going" to minimize human influence. A migration process was terminated under one of three conditions: the agent confirmed the migration was complete; the agent encountered an unrecoverable error; or the agent entered an infinite loop. 

\subsection{Prompt Engineering}

We engineered two distinct prompts to perform the migration. The first, shown in Figure \ref{fig:copilot-instructions-prompt}, was designed to generate the Copilot Instructions file that would guide the subsequent migration. Drawing insights from previous research \cite{automatic-library-migration-using-llms}, we provided the model with an example of an already migrated code snippet to improve the quality of the output. The prompt also included explicit directives, such as specifying the target library version and instructing the model to avoid introducing new functionality. The instructions file generated by this prompt is available in the paper's replication package.

\begin{figure}[!h]
  \centering \small
  \fcolorbox{gray!60}{lightergray}{%
    \parbox{\dimexpr\linewidth-2\fboxsep-2\fboxrule\relax}{%
      \noindent\makebox[\linewidth]{\colorbox{customgray}{\parbox{\dimexpr\linewidth-2\fboxsep-2\fboxrule\relax}{\textbf{\color{white}Copilot Instructions Creation Prompt}}}}\\[0.2cm]
      The Python code in this repository uses the library SQLAlchemy with version 1. We will migrate it so that it works with version 2 of SQLAlchemy. We must also make the code compatible with python’s \texttt{asyncio} and use python’s \texttt{typing} module to add type hints to the code. We must not add extra functionality to the code. Create a \texttt{migrate-sqlalchemy.instructions.md} file detailing the migration process and what should be done to achieve this migration. Use the example provided in the \texttt{sqlalchemy.py} file and the official documentation as reference to create the instructions file. The instructions file should only reference the SQLAlchemy migration, focusing on details needed to upgrade it between versions.
    }%
  }
  \caption{Prompt to generate the Copilot Instructions file}
  \label{fig:copilot-instructions-prompt}
\end{figure}

The second prompt, presented in Figure \ref{fig:migration-prompt}, was used to execute the actual migration. In this prompt, we directed the LLM agent to use the newly created \texttt{migrate-sqlalchemy.instructions.md} file as its primary guide for upgrading the repository to SQLAlchemy v2. This prompt was executed once for each of the 10 repositories in our dataset. To facilitate a realistic development workflow, we included instructions for managing the environment, such as using \texttt{uv} as the package manager and a local Python virtual environment. We also instructed Copilot to generate a TODO list to approach the task systematically. Finally, we provided credentials for a locally running PostgreSQL Docker container for repositories that required a database connection during test execution.

\begin{figure}[!h]
  \centering \small
  \fcolorbox{gray!60}{lightergray}{%
    \parbox{\dimexpr\linewidth-2\fboxsep-2\fboxrule\relax}{%
      \noindent\makebox[\linewidth]{\colorbox{customgray}{\parbox{\dimexpr\linewidth-2\fboxsep-2\fboxrule\relax}{\textbf{\color{white}Migration Prompt}}}}\\[0.2cm]
      Using the \texttt{migrate-sqlalchemy.instructions.md} instructions file, upgrade this repository so that it works with SQLAlchemy v2.\\
      Use \texttt{uv} as the python package and project manager\\
      Use Python's virtual environment for development and testing\\
      Before starting, define a TODO list with what you need to do and then systematically perform each task, without asking me for help.\\
      If you need to use a database connection, use the postgres docker container running locally with the following credentials\\
      username: postgres \\
      password: postgres \\
      database: <REPOSITORY\_NAME>
    }%
  }
  \caption{Prompt to migrate the repositories}
  \label{fig:migration-prompt}
\end{figure}

\subsection{Evaluation Metrics}
\label{sec:metrics-definition}

To assess the effectiveness of the migration and evaluate whether the migrated application functions as intended, we defined four categories of metrics, detailed in the following subsections. 

\subsubsection{Migration Coverage}

To measure the agent's effectiveness at the code-transformation level, we defined a Migration Coverage metric. The methodology is an adaptation of the work by \citet{characterizing-python-library-migrations}, which characterized API changes to understand library evolution.  We apply these principles to quantify the extent to which the LLM correctly migrated the API usage in a given client application.

To calculate this, we first manually identify all library usages requiring an update. We then use a table (see an example in Table \ref{tab:migration-coverage-example-table}) to systematically track each required transformation. The first two columns define the change (e.g., renaming \texttt{Column} to \texttt{mapped\_column}). The third column records the total number of instances where this change is needed, and the fourth column records the number of times the LLM successfully performed it. Migration Coverage is calculated as the ratio between the sum of the values in the fourth column and the sum of the values in the third column.

\begin{table}[!h]
    \centering
    \caption{Migration Coverage Example}
    \begin{tabular}{cccc}
        \toprule
        \textbf{Before} & \textbf{After} & \textbf{Instances} & \textbf{Score} \\
        \midrule
        \texttt{Column} & \texttt{mapped\_column} & 3 & 2 \\
        \texttt{create\_engine} & \texttt{create\_async\_engine} & 5 & 2 \\
        \texttt{session.query} & \texttt{select} & 3 & 1 \\
        \midrule
        \textbf{Total} & & 11 & 5 \\
        \bottomrule
    \end{tabular}
    \label{tab:migration-coverage-example-table}
\end{table}

An example is shown in Table \ref{tab:migration-coverage-example-table}, referring to an API with three changes. We can see that the class \texttt{Column} (first column) was renamed to \texttt{mapped\_column} (second column) in the new version of the API. In the system under analysis, this class is used in three source code locations (third column), of which two were correctly migrated by the LLM (fourth column). We can also see the old API was referenced in 11 code location, of which 5 were correctly migrated by the LLM. Thus, the Migration Coverage for this example is $5 / 11 \approx 45\%$. 

\subsubsection{Percentage of Passing Tests}
This metric evaluates the efectiveness of the migration by measuring the percentage of passing tests before and after the migration for all 10 repositories. 

\subsubsection{Application Compiles}

This metric assesses whether the application successfully compiles after migration. Both interpretation errors and static analysis issues are considered compilation problems. Examples include an \texttt{ImportError} (due to an invalid import statement) and \texttt{SyntaxError} (due to invalid coding syntax).

\subsubsection{Quality Metrics}

The last group of metrics are the quality metrics. These aim to assess the quality of produced code, checking whether the migrated code maintains the same level of quality when compared to the version before migration. Two metrics were used for this category. The first one is the Pylint score. This tool is a commonly used linter in Python. Thus, after the code was migrated, we executed the linter in order to check for common errors, such as missing imports or unused variables. 
The second metric is the number of Pyright errors. Pyright is a type checker tool for Python. It was used to identify typing errors before and after each migration.

\section{Results}
\label{sec:results}
In this section, we evaluate the library migration using Github's Copilot Agent Mode with GPT-4o as an LLM model. The results for each metric are detailed in Section \ref{sec:quantitave-assessment}. In Sections \ref{sec:discussion} and \ref{sec:previous-work}, we discuss our results and compare with a previous non-agentic approach, respectively. 

\subsection{Quantitative Assessment}
\label{sec:quantitave-assessment}

Table \ref{tab:migration-results} presents the results for the quantitative evaluation, using the methodology described in Section~\ref{sec:metrics-definition}. We also present measures for the code before the migration, as a baseline for comparison. In particular, both Migration Coverage and Passing Tests metrics are presented in two forms: \textit{aggregate} and \textit{median}. The aggregate results are obtained by analyzing all projects collectively, as if they constituted a single repository. In contrast, the median results are calculated across the individual outcomes of each of the ten studied repositories.

\begin{table}[htb]
    \centering
    \caption{Migration Results}
    \begin{tabular}{lcc}
        \toprule
        Metrics & \textbf{Before Migration} & \textbf{After Migration} \\
        \midrule
        \multicolumn{3}{l}{\textit{Migration Coverage}} \\
        \quad Aggregate (\%) & - & 45.48 \\
        \quad Median (\%) & - & 100 \\
        \midrule
        \multicolumn{3}{l}{\textit{Passing Tests}} \\
        \quad Aggregate (\%) & 87.84 & 53.61 \\
        \quad Median (\%) & 100 & 39.75 \\
        \midrule
        Compiling Repos (\#) & 10 & 8 \\
        Avg. Pylint Score & 6.16 & 6.48 \\
        Avg. Pyright Errors & 45.8 & 35.6 \\
        \bottomrule
    \end{tabular}
    \label{tab:migration-results}
\end{table}

\noindent\textbf{Migration Coverage.} The results for this metric, as presented in Table \ref{tab:migration-results}, show complementary insights. While the median has a value of 100\%, the aggregate measure shows that only 45.48\% of the code was actually migrated in the whole dataset. 
This was mostly due to the performance achieved in the \texttt{sqlalchemy-adapter}. This repository contains the highest number of SQLAlchemy usages in the dataset, with 124 instances that require migration. However, Copilot was able to migrate only 9.67\% of them. A key factor that hindered this process was a conflict with the \texttt{casbin} library. The \texttt{sqlalchemy-adapter} project uses version 0.8.1 of \texttt{casbin}, which lacks compatibility with Python's \texttt{asyncio}---a core requirement to the migration to SQLAlchemy version 2. 

\begin{figure}[htb]
  \centering \small
  \fcolorbox{gray!60}{lightergray}{%
    \parbox{\dimexpr\linewidth-2\fboxsep-2\fboxrule\relax}{%
      \noindent\makebox[\linewidth]{\colorbox{customgray}{\parbox{\dimexpr\linewidth-2\fboxsep-2\fboxrule\relax}{\textbf{\color{white}Copilot Chat Thought Process}}}}\\[0.2cm]
      The \texttt{casbin} package is installed, but it seems the \texttt{async\_adapter} module is not available in this version.
      This could be due to a version mismatch or a missing feature in the installed version of \texttt{casbin}. \\
      
      I will check the \texttt{casbin} documentation or source code to confirm the availability of the \texttt{async\_adapter} module. \\
      
      The \texttt{async\_adapter} module is not part of the \texttt{casbin} package itself but is provided by external adapters, such as the \texttt{casbin\_async\_sqlalchemy\_adapter}. This adapter is specifically designed for \texttt{async} operations with SQLAlchemy. \\
      
      I will install the \texttt{casbin\_async\_sqlalchemy\_adapter} package to resolve the missing import issue.
    }%
  }
  \caption{Agent reasoning on the \texttt{casbin} issue with \texttt{asyncio} for the \texttt{sqlalchemy-adapter} repository.}
  \label{fig:agent-casbin-asyncio-issue-thought-process}
\end{figure}

As shown in Figure \ref{fig:agent-casbin-asyncio-issue-thought-process}, the agent indeed identified this compatibility issue. However, it attempted to install the \texttt{casbin\_async\_sql\-alchemy\_adapter} package to fix the issue, but this created a new problem. The application under migration uses a class called \texttt{Async\-Adapter}, which is not present in the installed package. This resulted in a loop where the agent would repeatedly try to consult the library documentation, fail, and request manual intervention. In line with our methodology, we answered with "keep going" three times, and after observing the agent was in loop, we decided to abort the process. Interestingly, a simpler fix existed: updating the \texttt{casbin} library to version 1.23.0, which has \texttt{asyncio} support, and would have resolved the issue without the need for a new library.

\vspace{2mm}
\noindent\textbf{Passing Tests.} Analyzing the Passing Tests results, we observe that it achieved 39.75\% of success for the aggregate measures and a  higher value for the median (53.61\%). 
Out of the 10 migrated repositories, only two repositories had 100\% of passing tests after the migration.
Two other repositories experienced test failures due to assertion errors. For instance, the \texttt{paracelsus} repository includes a test that verifies whether converting a \texttt{Mermaid} type to a \texttt{string} produces the output \texttt{"True if post is published,nullable"}. However, the output was inverted after the migration, resulting in \texttt{"nullable,True if post is published"}. This indicates that the agent did, in fact, change the code’s behavior during the migration.

The remaining repositories showed failing tests due to syntax errors during the migration process. 
For instance, the \texttt{db\_to\_sqlite} repository faced migration issues related to the way the agent attempted to fix its tests.
We observed that 18 (out of 19) failing test cases were actually skipped using the \texttt{skipif} decorator. 
This represented an attempt by the agent to skip tests when a specific library was not installed. However, after manually installing the library, removing the decorator and rerunning the tests, they still failed.

\vspace{2mm}
\noindent\textbf{Compiling Repositories.} 
Out of the ten migrated repositories, eight successfully compiled after the migration. The two that failed were \texttt{nebulo} and \texttt{alembic\_utils}. The first failure occurred due to an incomplete migration: the agent stopped midway because of a local virtual environment issue, leaving the application in an unrecoverable state. The second failure was caused by a \texttt{CircularImport} error, which occurs when a module attempts to import another module that is only partially loaded.

\vspace{2mm}
\noindent\textbf{Quality Metrics.} Interestingly, the static analysis metrics suggest an improvement in code quality after the automated migration. The average Pylint score increased from 6.16 to 6.48, indicating that the LLM-generated code adhered more closely to standard Python style and conventions. More significantly, the average number of Pyright type errors decreased from 45.8 to 35.6. This indicates that the agent was able to correctly apply Python's \texttt{typing} module in the code, resulting in better type safety, which may reduce errors in future development.


\vspace{2mm}
\subsection{Discussion}
\label{sec:discussion}

An analysis of the migration results highlights an interesting pattern. First, a group of five repositories emerged, representing half of our dataset. These projects achieved perfect migration coverage (100\%) and high test pass rates (over 80\%, with two repositories having 100\%). One project, \texttt{FastApi-Strawberry-GraphQL-Sql\-Alchemy\-BoilerPlate}, stood out as a perfect migration case: it achieved both 100\% migration coverage and a 100\% test pass rate, along with an improved Pylint score (from 0 to 3.1) and reduced Pyright errors (from 39 to 32). This shows the agent’s capability to successfully complete migrations in particular cases.

The other five repositories had test pass rates below 80\%. However, this functional failure was not necessarily due to a ``lack'' in code migration, since three of these five projects had migration coverage scores of 80\% or higher. Thus, in these three projects, the agent understood and correctly identified the code requiring update but failed to update the codebase without breaking the application, leading to test failures.

\subsection{Comparison with a Non-Agentic Approach}

In a previous short paper~\cite{automatic-library-migration-using-llms}, we evaluated the use of GPT-4 for migrating SQLAlchemy from version 1 to version 2 in a single client application (\texttt{BiteStreams/fastapi-template}).
We used a tradtitional and non-agentic approach based on three  prompt strategies: Zero-Shot, One-Shot, and Chain-of-Thought. Table \ref{tab:one-shot-vs-agent-results} presents a comparison of the performance of the best prompt (One-Shot) against Copilot’s Agent Mode (for the single client used in our previous work and also included in the dataset of this new paper).

\begin{table}[htb]
    \centering
    \caption{One-Shot vs Copilot Agent Mode (\texttt{fastapi-template})}
    \begin{tabular}{lcc}
        \toprule
        Metrics & \textbf{One-Shot} & \textbf{Copilot Agent Mode} \\
        \midrule
        Migration Coverage (\%) & 81.25 & 100 \\
        Passing Tests (\%) & 25 & 50 \\
        Compiles (\#) & true & true \\
        Avg. Pylint Score & 7.77 & 7.93 \\
        Avg. Pyright Score & 23 & 13 \\
        \bottomrule
    \end{tabular}
    \label{tab:one-shot-vs-agent-results}
\end{table}

As we can see, there is a noticeable improvement when comparing both strategies, since Copilot's Agent Mode performed better in all four metrics. It obtained a migration coverage of 100\%, compared to 81.25\% previously observed. While only 25\% of the tests passed after the migration with the One-Shot prompt, 50\% passed with the newer strategy. Regarding quality metrics, Pylint registered a slight improvement (7.77 vs 7.93) and there was a relevant decrease in the number of errors using Copilot's Agent Mode (23 vs 13). In summary, these results indicate that using an agent may improve migration performance compared to a prompt-based approach.

\label{sec:previous-work}

\section{Threats to Validity}
\label{sec:threats-to-validity}

The findings in this paper are limited to the performance of GPT-4o. This means that using different LLMs, such as Google Gemini or Anthropic Claude, as well as different versions of GPT, might yield different results. Moreover, replicating this study with different agents would help validate our findings across a broader set of tools. We also restricted our study to a single Python library, which poses a threat to the external validity of our results. Thus, VSCode Copilot Agent’s performance in this specific ORM migration may not reflect its capabilities with other types of libraries. Furthermore, our findings should not be generalized to migrations in other programming languages without further investigation.

\section{Related Work}
\label{sec:related-work}


A recent large-scale study at Google demonstrated the use of Large Language Models (LLMs) to automate code migrations across dozens of production systems, where LLMs generated about 74\% of the required edits and reduced overall migration time by nearly half~\cite{migrating-code-at-scale-google}. Their approach followed a human-in-the-loop workflow, in which AI-generated patches were reviewed and validated by developers before integration. In contrast, our work investigates a fully autonomous agentic setup to perform end-to-end library migrations without human supervision. 

\citet{islam2025empiricalstudypythonlibrary} quantify the ability of large language models (LLMs) to perform library-migrations in Python by evaluating three LLMs (Llama 3.1, GPT-4o mini, GPT-4o) on the PyMigBench benchmark of 321 real-world migrations and 2,989 code changes. The authors report correct migration of ~89–94\% of the changes, and passing of the original unit tests in ~36–64\% of cases. They also extend evaluation to 10 unseen repositories to check for memorization. The study therefore provides empirical evidence that LLMs can effectively automate API-migration tasks, while also identifying remaining challenges in test-preservation and unseen-code generalization. In contrast, our work uses an agentic workflow that automates migrations (including code transformation and library upgrade) without human supervision.

\section{Conclusion and Future Work}
\label{sec:conclusion}

This study evaluated the effectiveness of Copilot's Agent Mode powered by GPT-4o in migrating ten repositories from SQLAlchemy v1 to v2. Our findings show that for eight of the ten repositories the LLM Agent achieved a migration coverage of over 80\%. Out of these, five repositories also had a passing test rate of over 80\%, demonstrating that the agent is capable of producing a complete migration requiring little to no manual intervention. However,the remaining five repositories had test pass rates below 80\%. These results, when analyzed together with the migration coverage results, suggest that while the agent understands which API migrations are required, it sometimes struggles to preserve overall application functionality, resulting in test failures.

Based on our findings, a follow-up study should investigate a human-in-the-loop approach. Instead of using a passive ``keep going'' response, an experiment could be designed where a developer actively interacts with the agent, providing feedback and correcting errors. Another future work may explore methods to provide the agent with a richer understanding of a project's runtime behavior. An example would be to have specific instructions files detailing the libraries used in the project and how they should be executed.

\vspace{2mm}
\noindent{\bf Replication Data:} Scripts, prompts and results of this research are available at: \url{https://zenodo.org/records/17364057}

\begin{acks}
This research was supported by grants from CNPq and FAPEMIG.
\end{acks}

\bibliographystyle{ACM-Reference-Format}
\bibliography{references}

\end{document}